\documentclass[preprint]{aastex}
\usepackage{amssymb,amsmath,epsfig}

\newcommand{\ms}{\mbox{m s$^{-1}$}}
\newcommand{\cms}{\mbox{cm s$^{-1}$}}

\newcommand{\mylabelf}[1]{\label{fig_#1}}
\newcommand{\mylabele}[1]{\label{eq_#1}}

\newcommand{\mylabelt}[1]{\label{tab_#1}}

\newcommand{\fig}[1]{Figure~\ref{fig_#1}}

\newcommand{\seefig}[1]{(see Figure~\ref{fig_#1})}
\newcommand{\equ}[1]{Eq.~(\ref{eq_#1})}

\newcommand{\tabl}[1]{Table~\ref{tab_#1}}

\newcommand{\br}[1]{\left(#1\right)}

\shorttitle{Double fiber scrambler at Keck Observatory} 
\shortauthors{J. Spronck et al}

\begin{document} 

\title{Fiber scrambling for high-resolution spectrographs. II. A double fiber scrambler for Keck Observatory}

\author{Julien F.P. Spronck\altaffilmark{1},
    Debra A. Fischer\altaffilmark{2},
    Zachary Kaplan\altaffilmark{2}, 
    Colby A. Jurgenson\altaffilmark{2}, 
    Jeff Valenti\altaffilmark{3},
    John Moriarty\altaffilmark{2}, 
    Andrew E. Szymkowiak\altaffilmark{2}}

\altaffiltext{1}{Leiden Observatory, Leiden University, 
    Niels Bohrweg 2, 2333 CA, Leiden, The Netherlands}
\altaffiltext{2}{Department of Astronomy, 
    Yale University, New Haven, CT 06511, USA}
\altaffiltext{3}{Space Telescope Science Institute,
       3700 San Martin Dr., Baltimore, MD 21218}

\email{spronck@strw.leidenuniv.nl, julspronck@gmail.com}

\begin{abstract}
We have designed a fiber scrambler as a prototype for the Keck HIRES spectrograph, using double scrambling to stabilize illumination of the spectrometer and a pupil slicer to increase spectral resolution to R$\sim$70,000 with minimal slit losses. We find that the spectral line spread function (SLSF) for the double scrambler observations is 18 times more stable than the SLSF for comparable slit observations and 9 times more stable than the SLSF for a single fiber scrambler that we tested in 2010. For the double scrambler test data, we further reduced the radial velocity scatter from an average of $\sim$2.1 \ms\ to $\sim$1.5 \ms\ after adopting a median description of the stabilized SLSF in our Doppler model. This demonstrates that inaccuracies in modeling the SLSF contribute to the velocity RMS. Imperfect knowledge of the SLSF, rather than stellar jitter, sets the precision floor for chromospherically quiet stars analyzed with the iodine technique using Keck HIRES and other slit-fed spectrometers. It is increasingly common practice for astronomers to scale stellar noise in quadrature with formal errors such that their Keplerian model yields a chi-squared fit of 1.0. When this is done, errors from inaccurate modeling of the SLSF (and perhaps from other sources) are attributed to the star and the floor of the stellar noise is overestimated. 
\end{abstract}


\section{Introduction}
Steady improvements in Doppler instrumentation and analysis techniques \citep{Marcy1992, Butler1996, Mayor2003} over the past twenty years have produced two orders of magnitude improvement in measurement precision, enabling the detection of hundreds of exoplanets \citep{Latham1989, MayorQueloz95, Pepe2011, Fischer2014}. However, transit observations by the European Space Agency {\it CoRoT} spacecraft and the NASA {\it Kepler} mission have shown that small rocky planets are far more common \citep{Fressin2013, Dressing2013, Howard2012, Buchhave2014} than the population of more massive planets detected with the Doppler technique. While space-based transit techniques discover and measure the sizes of small planets, complementary Doppler measurements are required to derive the key attributes of mass and density. In order to be scientifically competitive in the future, Doppler precision must be improved by another order of magnitude to 10 \cms. The required improvement in Doppler precision is technically challenging. It is not simply a matter of building a more stable spectrometer. Thoughtful innovations will be required, including higher spectral resolution to model and decorrelate stellar photospheric noise, new wavelength calibrators \citep{Langellier2014}, and optimal coupling of light into the spectrometer \citep{Spronck_2012_SPIE_2}.

Optical fibers offer simple mechanical design and have been used since the 1980’s to transport light from the telescope focus to the entrance of high-precision spectrometers. However, optical fibers have an additional useful attribute: they scramble light \citep{Heacox1980, Heacox1986, Heacox1988, Barden1981}, producing an output beam that is nearly independent of the input. The scrambling property of optical fibers stabilizes illumination of the spectrometer optics and reduces spurious spectral line profile variations compared to slit-fed spectrometers. The improved stability of the spectral line profile that is possible with fiber coupling is critical for improving the Doppler measurement precision. 

Our current Doppler technique uses a gas cell filled with molecular iodine to imprint thousands of reference spectral lines onto the observed stellar spectrum \citep{Marcy1992, Butler1996}. The iodine lines serve as a wavelength calibrator and are used to model changes in the instrumental profile, also known as Spectral Line Spread Function (SLSF). Contributions to the SLSF come from both the spectrometer optics and from the illumination of the slit and pupil. The SLSF contribution from the spectrometer varies slowly over time as temperature and pressure changes occur within the spectrometer. These slow SLSF variations are easy to track and remove in our Doppler model. However, the SLSF contributions from illumination of a slit-fed spectrometer vary on a timescale of seconds as a result of guiding errors or changes in the focus or seeing. SLSF variations from one integration to the next contribute significantly to the total Doppler error budget because modeling of the SLSF is inaccurate. Optical fibers are promising because they attenuate these rapid variations in the SLSF, although their scrambling capability is not perfect and the fiber output can still retain some dependence on the way light is injected \citep{Spronck_2013_PASP}. The goal of fiber coupling of light into spectrometers is to minimize variability in the spectrometer input. 

Early tests showed that optical fibers could improve the stability of the SLSF \citep{Spronck_2013_PASP}, even on instruments like the Hamilton spectrograph \citep{Vogt1987} that were not engineered for extreme stability. This provided motivation to develop a first prototype fiber scrambler in 2010 \citep{Spronck_2012_SPIE_1} for the Keck HIRES spectrograph \citep{Vogt1994}. Physical space constraints were tight with less than 2 inches available to pick off the light between the iodine cell and the decker plate. Re-imaging optics at the entrance of the fiber were used to convert the f-ratio to minimize focal ratio degradation. A 10-m fiber with a circular cross-section was used to scramble the light. Re-imaging optics after the fiber converted back to the focal ratio of the Keck telescope before re-injecting the beam into HIRES. This prototype significantly improved the stability of the SLSF; however, we were unable to see any improvement in our real figure of merit, the radial velocity precision \citep{Spronck_2012_SPIE_1}. Subsequent analysis suggested that while we had improved the stability of the SLSF, the error budget for the Doppler precision was limited by other problems, including a loss in resolution with the circular fiber and lower signal-to-noise ratio, probably due to the focal ratio degradation overfilling the collimator in HIRES. 

\section{The second prototype: a double scrambler design}

The single circular fiber scrambler \citep{Spronck_2012_SPIE_1} is very good at scrambling the light in the azimuthal direction but not as good in the radial direction. In order to stabilize intensity distributions in both azimuthal and radial directions, we need to scramble both the near-field (the light distribution at the end of the fiber) and the far-field (the light distribution in the pupil plane) distributions. A double scrambler does just that by
re-imaging the far-field of a first fiber onto the input face of a second fiber, thereby inverting and scrambling both far-field and near-field \citep{Hunter1992,Avila1998}.

We had two goals when redesigning the fiber scrambler prototype:  increasing spectral resolution relative to the circular fiber and achieving more complete scrambling with octagonal, square or rectangular fibers, which are more efficient at mode mixing than traditional circular fibers \citep{Chazelas2010, Spronck_2012_SPIE_2, Plavchan2013}. Based on preliminary work done at Lick Observatory \citep{Spronck_2013_PASP}, we also wanted to implement double scrambling. Double fiber scramblers are commonly built with microscopic ball or rod lenses. While this has worked successfully in the past, these systems are relatively difficult to build and align, therefore leading to low throughput \citep{Spronck_2013_PASP}. Additionally, these systems can suffer chromaticity. We therefore decided to build a macroscopic double scrambler \citep{Barnes2010} composed of two achromatic doublets.

\subsection{System overview}

The overall system is depicted in \fig{system}. Light from the telescope is first re-directed by a pick-off mirror to a 640-micron (0.95 arcsec) pinhole in a tilted plate located in the telescope focal plane. The surface of the aperture plate is mirrored, so that low intensity wings of the star image outside the pinhole reflects back to the guider camera. The 25-mm apertude plate has a field of view of 37~arcsec and is held using a gimbal mount to ensure that the pinhole remains at the telescope focus when manually adjusting tip/tilt of the plate (see \fig{system} and \fig{fembem}).

\begin{figure}[ht]
\epsscale{0.75}
\plotone{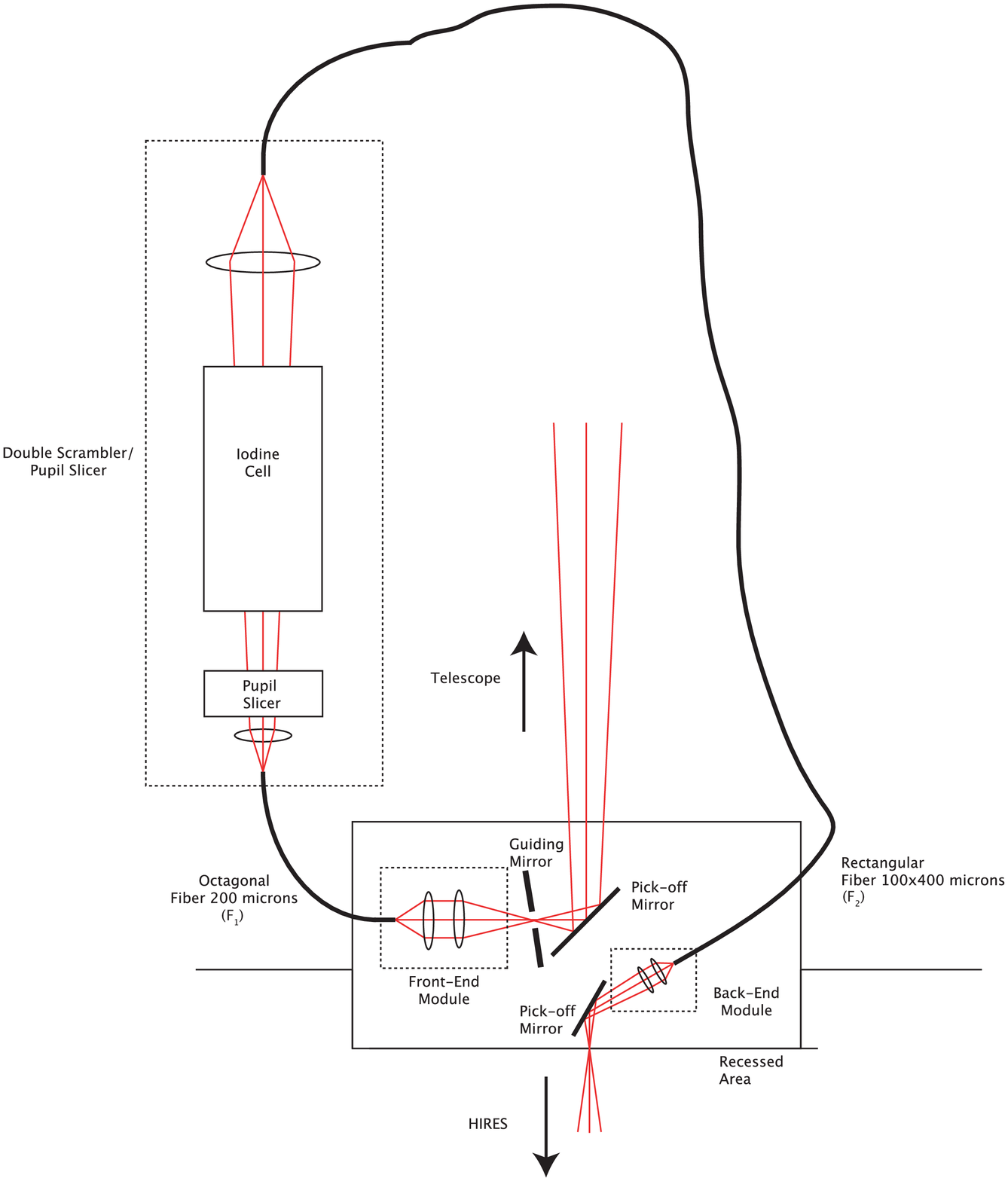}
\caption{Schematic view of the entire system: pick-off mirrors, front-end module, back-end module, double scrambler and pupil slicer. Each of the six lenses in the figure is an achromatic doublet.}
\mylabelf{system}
\end{figure}

The beam that passes through the pinhole is then sent into an octagonal fiber ($F_1$) by a pair of achromatic doublets (see \fig{system}) fixed to a common monolithic barrel on the front-end module. This barrel centers the lenses with a fixed predetermined separation. The fiber plugs into a standard FC/PC connector and both the lens barrel and the fiber have lateral and longitudinal adjustability for optimizing the throughput of the system.

The octagonal fiber scrambles the near-field and the beam that emerges then passes through a series of elements: the achromatic doublets that ensure double scrambling, the pupil slicer and the iodine cell. The sliced pupil is imaged onto the face of a rectangular fiber ($F_2$). The rectangular fiber scrambles the far-field and the fiber output is re-imaged with a second pair of achromatic doublets (on the back-end module) at the position where the star would have been focused during normal operations (without a fiber scrambler) and sent towards HIRES with another mirror. The main components of the system are discussed in more detail below.

\begin{figure}[ht]
\epsscale{0.65}
\plotone{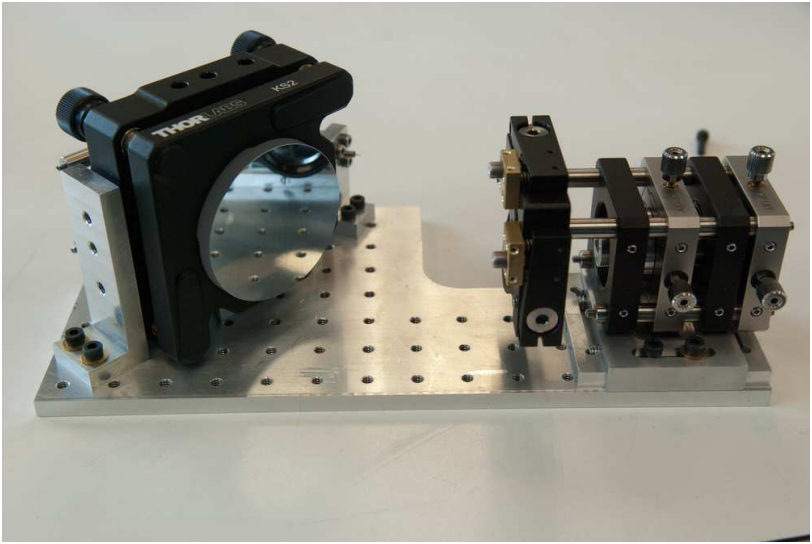}
\caption{Picture of the breadboard containing the front-end and the back-end modules as seen by the incoming beam. To the left is the mirror that picks-off the light from the telescope and to the right is the front-end module with the guiding mirror, the two doublets and the fiber holder. The back-end module (not visible on the picture) is behind the pick-off mirror.}
\mylabelf{fembem}
\end{figure}

\subsection{Optical fibers}
In order to further improve the stability of the double scrambler, we used non-circular fibers that are inherently better scramblers than circular fibers \citep{Chazelas2010,Spronck_2012_SPIE_2,Plavchan2013}: The first fiber ($F_1$) is octagonal while the second ($F_2$) is rectangular. Both fibers were made by Ceramoptec with an acrylate jacket to prevent light from propagating through the cladding of the fiber. The octagonal fiber was 200~$\mu$m from flat to flat, while the rectangular fiber was $100 \times 400 \mu$m.

\subsection{The macroscopic double scrambler}
The macroscopic double scrambler is composed of two achromatic doublets \seefig{ds}. 
The first doublet ($L_1$) collimates the beam coming out of the first (octagonal) fiber ($F_1$), and the second doublet ($L_2$) re-images the pupil onto the second (rectangular) fiber ($F_2$). The design parameters are $f_1$, $f_2$ and $d$, where $f_1$ and $f_2$ are the focal lengths of $L_1$ and $L_2$, and $d$ is the distance between the two doublets.

\begin{figure}[htbp]
\epsscale{0.85}
\plotone{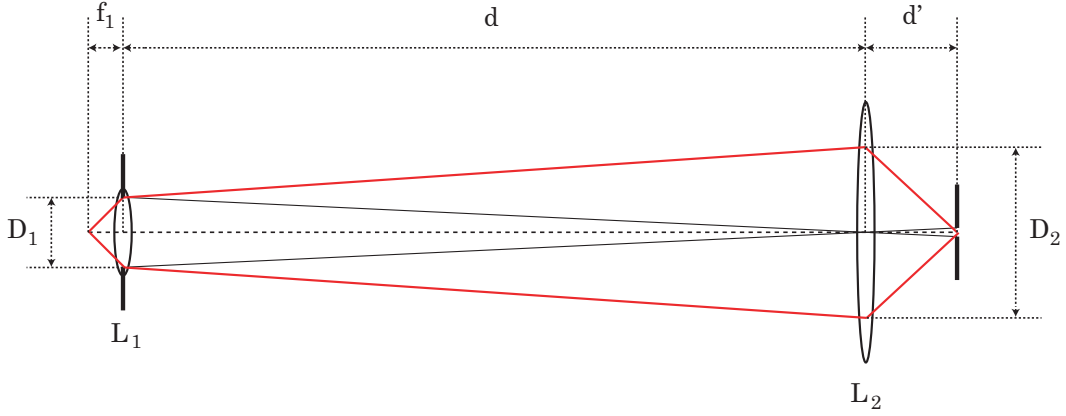}
\caption{Schematic drawing of the macroscopic double scrambler.}
\mylabelf{ds}
\end{figure}

The diameter of the fiber ($d_f$) is set by the telescope diameter ($D_t$), the focal ratio of the beam entering $F_1$ ($F_{\#,1}$) and the desired angular size of the fiber on sky ($\theta$):
\begin{equation}
\mylabele{fibsize}
d_f \approx 5 * D_t * F_{\#,1} * \theta,
\end{equation}
where $d_f$ is in microns, $D_t$ is in meters and $\theta$ is in arcseconds. The factor 5 in \equ{fibsize} comes from the conversion from meters and radians to microns and arcseconds. For a 10-m telescope at F/4, a 1"-fiber has a diameter $d_f = 200$~microns.

The pupil size on $L_1$ is denoted by $D_1$. In order to re-image the pupil on $F_2$, we need to de-magnify it by a factor $M = D_1/d_f$. From geometrical optics, we know that the object distance (as far as the imaging system formed by $L_2$ is concerned) $d$ is given by
\begin{equation}
d = f_2 (1+M).
\mylabele{d}
\end{equation}

Another parameter that comes into play in the design of the double scrambler is focal ratio degradation (FRD). Not only does each fiber introduce FRD ($\mbox{FRD}_1$ and $\mbox{FRD}_2$), but the double scrambler itself will introduce FRD that should be minimized. In order to do that, the focal ratio of the beam entering fiber $F_2$ ($F_{\#,2}$) should be as close as possible to the focal ratio exiting the fiber $F_1$ ($F_{\#,1}*\mbox{FRD}_1$), 
\begin{equation}
F_{\#,2} = \frac{d'}{D_2},
\mylabele{F2}
\end{equation}
where  $d'$ is the image distance and $D_2$ is the beam size on $L_2$. From geometrical optics, we have
\begin{equation}
d' = f_2 \frac{1+M}{M}
\mylabele{dp}
\end{equation}
and
\begin{equation}
D_2 = D_1 + d * d_f / f_1.
\mylabele{D2}
\end{equation}

After inserting  \equ{dp}, \equ{d} and \equ{D2} in \equ{F2}, we find
\begin{equation}
F_{\#,2} = \frac{f_2(1+M)/M}{D_1+f_2(1+M)d_f/f_1} = \frac{f_2(1+M)/M}{D_1 \br{1+f_2(1+M)/M/f_1}}.
\mylabele{F22}
\end{equation}
$F_{\#,2}$ will be maximal when $f_2 >> f_1$. In that case, we have
\begin{equation}
F_{\#,2,max} \approx \frac{f_1}{D_1} = F_{\#,1}*\mbox{FRD}_1.
\mylabele{F23}
\end{equation}

This shows that the double scrambler introduces focal ratio degradation, which is minimized when $f_2 >> f_1$. The overall size of the system (the distances and the diameter of the optics) increases with $f_1$ and with $f_2$. We therefore need to make a trade-off between FRD and system size. As a compromise, we chose $f_1 = 6.25$~mm and $f_2 = 75$~mm. Commercial achromatic doublets for both these focal lengths are standard off the shelf optics. Using these focal lengths, the beam diameters are $D_1 = 1.6$~mm and $D_2 = 25$~mm and the overall system length is 750~mm.

\subsection{The pupil slicer}

In order to recover the resolution that we had lost with the previous (circular) fiber scrambler, we make use of a so-called pupil slicer. The slicer is a modified Bowen-Walraven slicer composed of two quasi-parallel half mirrors, similar to the slicer built for the CHIRON spectrograph \citep{Tokovinin2013} (\fig{slicer}). The two half mirrors were made by cutting one single mirror in two with a diamond tip in order to have very sharp edges. The slicer transforms the round pupil emerging from the first fiber into two half-moons and stacks these images on top of each other; the image after the slicer has an aspect ratio of 1 by 4 (\fig{slicer}). The stacked half-moons are then re-imaged by $L_2$ onto a rectangular $100 \times 400$-micron fiber while keeping the same focal ratio $F_{\#,2}$, which should effectively double the resolution.

\begin{figure}[ht]
\epsscale{0.75}
\plottwo{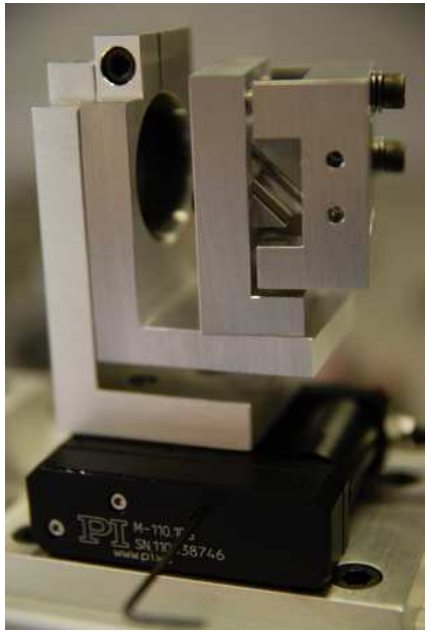}{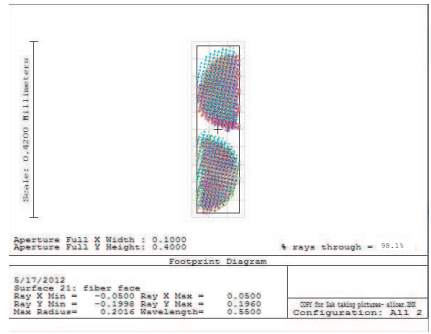}
\caption{Left: pupil slicer used to slice the pupil into two half-moons. With this design, all angles as well as the distance between the two mirrors are adjustable. Right: Slices as simulated by Zemax.}
\mylabelf{slicer}
\end{figure}

\subsection{Integration}

Since our prototype tests only lasted two nights, we had to install our scrambler while minimizing disturbance to the rest of the instrument. We used an approach similar to our previous test in 2010: we created a custom plate to support the small front-end and back-end breadboard. This plate has the same dimensions as the standard decker plate for HIRES and was swapped into the position of the decker plate. We disconnected the decker motor to make sure that the plate did not accidentally get a signal to move during our observation program. The space constraints severely restricted the size of the front-end and back-end module and made the mechanical design challenging. Normally, the iodine cell (used for wavelength calibration) is positioned 20-30~mm in front of the decker plate (the decker being tilted with respect to the beam to send light towards the guiding camera). We decided to gain more room for our front-end and back-end modules by locking the HIRES iodine cell in the ``out'' position (we also disconnected the iodine cell motor to avoid catastrophic failures). Since an iodine cell was needed for the observations, we constructed a new iodine cell and placed it in the diverging beam of the double scrambler, after the octagonal fiber. A small motorized stage moves the cell in and out of the beam as needed.

\section{SLSF stability and throughput}

On 2012 November 28 and 29, we tested the double scrambler on Keck HIRES by obtaining several consecutive observations of seven stars that have a long history of low RMS velocity. In the case of 55 Cnc, we distributed observations through the night to see if we could detect 55 Cnc e, the short period planet detected in transit by the MOST satellite \citep{Winn2011}. We also obtained 56 spectra of B-type stars; these hot rapidly rotating stars have featureless spectra and are ideal light sources for illuminating the iodine absorption cell to model the wavelength solution and the SLSF. The forward modeling technique \citep{Marcy1992, Butler1996, Fischer2014} begins with a R$\sim$1,000,000 FTS scan of the I2 cell and convolves this spectrum with the model SLSF, as described by \citet{Valenti1995}. The code employs a Levenberg-Marquardt algorithm to fit every 2-\AA\ chunk of the iodine line spectra. The B-star observations were distributed throughout both nights so that we could evaluate the stability of the SLSF for the double fiber scrambler.

We also wanted to compare the SLSF models of the iodine spectra obtained with double scrambler with the SLSF models for the circular fiber and the slit. The spectral line spread functions for these three coupling techniques are systematically different, but all can be reasonably approximated by a Gaussian, fitted to the central component of the SLSF. We adopted the full width half max (FWHM) of the fitted Gaussian as the metric for comparing slit, circular fiber and double scrambler SLSF models. In \fig{slsf}, the blue triangles (left) show the mean FWHM of a Gaussian fit to the SLSF model for all spectrum chunks in a set of slit-fed iodine observations, obtained over three nights with similar sky conditions. The green squares show the mean FWHM fit to the SLSF model for iodine spectra obtained in 2010 with the circular fiber \citep{Spronck_2012_SPIE_1} and the red crosses (right) show the FWHM of fits to the SLSF for iodine observations obtained over two consecutive nights with the double scrambler. 

The FWHM of the fitted Gaussian is measured in pixels. The vertical location of the symbols in \fig{slsf} is an indication of the relative spectral resolution, while the scatter in the points is a measure of variability in the SLSF. Relative to observations obtained with the slit (R=55,000; 3.3 pixels FWHM) the circular fiber scrambler observations were degraded in spectral resolution (R=40000; 4.9 pixels FWHM) while the spectra obtained with the double scrambler had a higher spectral resolution (R=70000; 2.8 pixels FWHM). We clearly see night-to-night SLSF variations for the slit data, while the double scrambler SLSF remains stable over both nights in our observing run. Furthermore, the FWHM corresponding to the double scrambler shows much less scatter, or equivalently, greater stability. The standard deviation of the FWHM fit to the SLSF model for the double scrambler is 18 times less than for the slit-fed spectra and 9 times smaller than for the circular fiber data obtained in 2010. Thus, the double fiber scrambler met the goals of increasing resolution relative to the circular fiber and improving scrambling to produce a more stable SLSF.

\begin{figure}[ht]
\epsscale{0.75}
\plotone{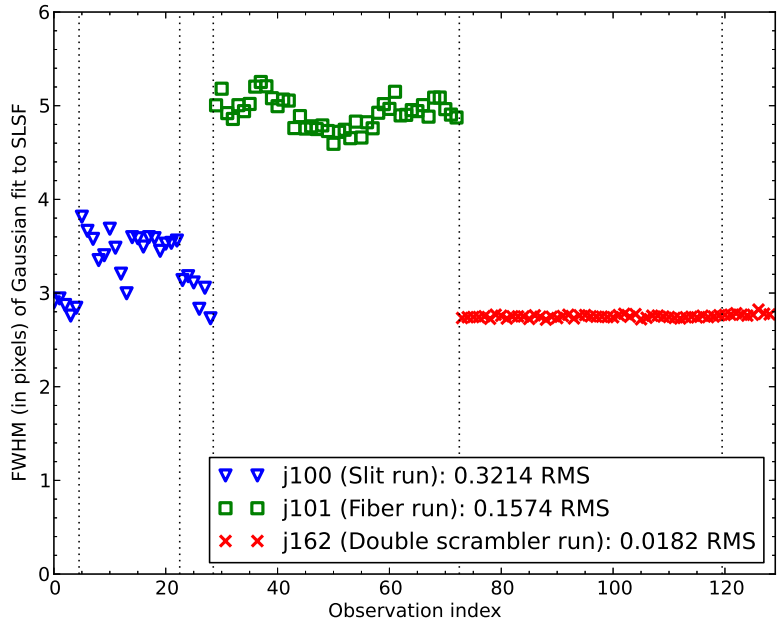}
\caption{Mean FWHM (in pixels) of a Gaussian fit to SLSF model of B-star-illuminated iodine spectra taken with (left panel, blue triangles) slit-fed observations obtained with the 0.861 arcsec B5 decker, (middle panel, green squares) a single circular fiber scrambler, and (right panel, red crosses) the double scrambler.The vertical dotted lines separate different observing nights.}
\mylabelf{slsf}
\end{figure}

Unfortunately, our on-sky assessment of the throughput for the double fiber system was considerably lower than what we measured when designing this system in the lab. We calculated the photons per second that were acquired with the slit and the double scrambler and found that the flux rate through the double scrambler was only 15\% of the flux rate for slit observations with similar sky conditions. We reexamined the double scrambler in the lab and measured 64\% throughput using a bright white LED; we concluded that the low throughput at the telescope was mainly due to an undetected misalignment at the telescope and to the focal ratio degradation of the system.

\section{Radial velocities}

The radial velocity single measurement precision (SMP) is a function of signal-to-noise ratio (SNR), analysis technique and spectral resolution. This metric reflects the ultimate floor of the Doppler precision for a given instrument and characterizes the instrumental precision. The SMP is generally free of stellar variability, which occurs on timescales that are longer than a single observation. The uncertainty for a single measurement in our analysis is the standard deviation of the chunk velocities after rejecting fewer than 1\% of the chunks with spurious velocities. The short-term velocity RMS (over several consecutive data points) is a better metric for characterizing instrumental precision because the time baseline is long enough to reveal the presence of instrumental and analysis errors but still short compared to changes in fractional spot coverage or stellar magnetic fields for chromospherically inactive stars.

In our Doppler analysis, we first generate a template observation for the forward modeling of our program observations. The template observation consists of 5 co-added spectra (shifted to a common velocity) that were obtained without the iodine cell and deconvolved with a spectral line spread function (SLSF) derived from adjacent B-star observations. The deconvolved template spectrum was multiplied by the high-resolution FTS scan of our iodine cell and convolved with an SLSF description to model the program observations of stars taken with the iodine cell. This technique models all 2-\AA\ chunks of the spectrum between 5010 - 6300 \AA, where there are iodine lines. The free parameters for each chunk include the wavelength zero point, the dispersion, the continuum offset and the Doppler shift, plus 17 free parameters for the SLSF description \citep{Valenti1995}. Given the dramatic improvement that we documented for the SLSF stability, we were surprised that we did not see an improvement in the radial velocity precision when we analyzed the data obtained with our double scrambler in this standard way.

\begin{deluxetable}{llcc}
\tablecaption{Radial Velocities from Double Scrambler}
\tablewidth{0pt}
\tablehead{ \colhead{ } & \colhead{JD-2440000} & \colhead{RV} & \colhead{Uncertainties}  \\
  \colhead{Star} & \colhead{[BJD]} & \colhead{(\ms)} & \colhead{(\ms)}    \\}
\startdata
HD~26965 &   16260.837891 &   0.15 & 0.89 \\
HD~26965 &   16260.838867 &  -1.77 & 0.93 \\
HD~26965 &   16260.839844 &   1.05 & 0.89 \\
HD~26965 &   16260.839844 &  -1.49 & 0.92 \\
HD~26965 &   16260.840820 &  -1.91 & 0.97 \\
HD~26965 &   16260.841797 &  -1.92 & 0.82 \\
HD~26965 &   16260.842773 &  -3.39 & 0.86 \\
HD~26965 &   16260.843750 &  -0.59 & 0.93 \\
HD~26965 &   16260.844727 &  -0.42 & 1.00 \\
HD~26965 &   16260.845703 &   0.35 & 0.90 \\
HD~26965 &   16260.846680 &  -2.65 & 0.85 \\
HD~26965 &   16260.847656 &   0.23 & 0.93 \\
HD~26965 &   16260.848633 &   1.23 & 0.90 \\
HD~26965 &   16260.848633 &  -2.96 & 0.79 \\
HD~26965 &   16260.849609 &   2.89 & 0.90 \\
HD~26965 &   16260.850586 &   0.06 & 0.99 \\
HD~26965 &   16260.851562 &   0.63 & 0.94 \\
HD~26965 &   16260.852539 &  -0.08 & 0.92 \\
HD~26965 &   16260.853516 &   1.19 & 1.08 \\
HD~26965 &   16260.854492 &   1.86 & 0.91 \\
HD~26965 &   16260.855469 &   0.75 & 0.92 \\
HD~26965 &   16260.856445 &  -0.55 & 0.95 \\
HD~26965 &   16260.857422 &   0.53 & 0.94 \\
HD~26965 &   16260.857422 &   1.78 & 0.95 \\
HD~26965 &   16260.858398 &  -2.81 & 0.91 \\
HD~10700 &   16260.782227 &  -0.53 & 1.05 \\
HD~10700 &   16260.783203 &   1.56 & 0.96 \\
HD~10700 &   16260.784180 &  -0.21 & 1.03 \\
HD~10700 &   16260.785156 &   1.67 & 0.96 \\
HD~10700 &   16260.785156 &   0.18 & 1.00 \\
HD~10700 &   16260.786133 &  -1.41 & 1.00 \\
HD~10700 &   16260.787109 &   0.83 & 1.10 \\
HD~10700 &   16260.788086 &  -1.86 & 1.07 \\
HD~10700 &   16260.789062 &   0.27 & 0.94 \\
HD~10700 &   16260.790039 &   1.07 & 1.03 \\
HD~10700 &   16260.790039 &  -0.22 & 1.10 \\
HD~10700 &   16260.791016 &   3.03 & 0.95 \\
HD~10700 &   16260.791992 &  -0.47 & 1.04 \\
HD~10700 &   16260.791992 &  -1.89 & 0.99 \\
HD~10700 &   16260.792969 &  -4.69 & 0.96 \\
HD~10700 &   16260.793945 &   0.72 & 0.94 \\
HD~10700 &   16260.794922 &  -1.07 & 0.88 \\
HD~10700 &   16260.794922 &  -1.59 & 0.93 \\
HD~10700 &   16260.795898 &  -0.57 & 0.87 \\
HD~10700 &   16260.796875 &  -0.01 & 0.99 \\
HD~10700 &   16260.798828 &  -2.29 & 2.29 \\
HD~10700 &   16260.799805 &   1.70 & 1.99 \\
HD~10700 &   16260.799805 &  -2.84 & 2.12 \\
HD~10700 &   16260.800781 &   2.55 & 1.88 \\
HD~10700 &   16260.800781 &  -3.08 & 2.48 \\
HD~10700 &   16260.801758 &   0.86 & 1.17 \\
HD~10700 &   16260.802734 &  -2.82 & 1.21 \\
HD~10700 &   16260.802734 &  -0.84 & 1.13 \\
HD~10700 &   16260.803711 &  -2.09 & 1.30 \\
HD~10700 &   16260.804688 &  -2.87 & 1.15 \\
HD~10700 &   16260.804688 &   2.66 & 0.83 \\
HD~10700 &   16260.805664 &  -1.75 & 0.84 \\
HD~10700 &   16260.806641 &  -1.48 & 0.76 \\
HD~10700 &   16260.807617 &  -0.88 & 0.80 \\
HD~10700 &   16260.808594 &  -1.04 & 0.78 \\
HD~10700 &   16260.809570 &   0.18 & 0.73 \\
HD~10700 &   16260.811523 &  -1.54 & 0.74 \\
HD~10700 &   16260.812500 &   0.29 & 0.73 \\
HD~10700 &   16260.813477 &  -0.19 & 0.69 \\
HD~10700 &   16260.777344 &   0.80 & 0.99 \\
HD~10700 &   16260.778320 &   0.64 & 1.11 \\
HD~10700 &   16260.779297 &   1.60 & 1.02 \\
HD~10700 &   16260.780273 &  -0.66 & 1.07 \\
HD~10700 &   16260.781250 &   0.90 & 0.96 \\
 HD~9407 &   16260.695312 &  -1.66 & 0.84 \\
 HD~9407 &   16260.700195 &  -0.64 & 0.80 \\
 HD~9407 &   16260.704102 &  -2.45 & 0.92 \\
 HD~9407 &   16260.708984 &  -2.64 & 0.84 \\
 HD~9407 &   16260.713867 &   0.86 & 0.83 \\
 HD~9407 &   16260.719727 &  -0.24 & 0.87 \\
 HD~9407 &   16260.724609 &   4.23 & 0.78 \\
 HD~9407 &   16260.728516 &   0.19 & 0.80 \\
 HD~9407 &   16260.732422 &   2.77 & 0.80 \\
 HD~9407 &   16260.736328 &  -0.43 & 0.83 \\
HD~32147 &   16260.883789 &  -0.61 & 0.77 \\
HD~32147 &   16260.886719 &   0.79 & 0.77 \\
HD~32147 &   16260.889648 &  -0.25 & 0.73 \\
HD~32147 &   16260.891602 &   1.98 & 0.69 \\
HD~32147 &   16260.894531 &   2.88 & 0.74 \\
HD~32147 &   16260.896484 &  -0.87 & 0.71 \\
HD~32147 &   16260.899414 &  -0.72 & 0.75 \\
HD~32147 &   16260.902344 &   1.10 & 0.73 \\
HD~32147 &   16260.905273 &   0.05 & 0.73 \\
HD~32147 &   16260.907227 &  -0.84 & 0.73 \\
HD~32147 &   16260.910156 &  -0.19 & 0.73 \\
HD~32147 &   16260.913086 &   0.45 & 0.75 \\
HD~32147 &   16260.915039 &   0.52 & 0.72 \\
HD~32147 &   16260.917969 &   0.23 & 0.70 \\
HD~32147 &   16260.919922 &   1.10 & 0.68 \\
HD~32147 &   16260.921875 &   1.40 & 0.68 \\
HD~32147 &   16260.923828 &   2.25 & 0.70 \\
HD~32147 &   16260.926758 &   2.00 & 0.77 \\
HD~32147 &   16260.928711 &  -0.01 & 0.79 \\
HD~32147 &   16260.930664 &   1.31 & 0.64 \\
HD~32147 &   16260.933594 &  -0.64 & 0.73 \\
HD~32147 &   16260.935547 &  -0.48 & 0.74 \\
HD~32147 &   16260.937500 &  -0.36 & 0.70 \\
HD~32147 &   16260.939453 &  -0.21 & 0.71 \\
HD~32147 &   16260.941406 &  -1.61 & 0.72 \\
HD~75732 &   16260.944336 &   5.01 & 0.80 \\
HD~75732 &   16260.947266 &   3.60 & 0.83 \\
HD~75732 &   16260.950195 &   3.48 & 0.87 \\
HD~75732 &   16260.953125 &   2.28 & 0.85 \\
HD~75732 &   16260.956055 &   4.11 & 0.74 \\
HD~75732 &   16261.005859 &   2.89 & 0.76 \\
HD~75732 &   16261.007812 &   3.52 & 0.78 \\
HD~75732 &   16261.010742 &   2.14 & 0.76 \\
HD~75732 &   16261.013672 &   1.43 & 0.79 \\
HD~75732 &   16261.015625 &   2.33 & 0.77 \\
HD~75732 &   16261.084961 &   1.62 & 0.82 \\
HD~75732 &   16261.086914 &  -4.04 & 0.78 \\
HD~75732 &   16261.089844 &   0.15 & 0.77 \\
HD~75732 &   16261.091797 &  -0.80 & 0.78 \\
HD~75732 &   16261.094727 &   1.12 & 0.80 \\
HD~75732 &   16261.155273 &  -3.97 & 0.77 \\
HD~75732 &   16261.159180 &  -4.87 & 0.79 \\
HD~75732 &   16262.013672 &  -1.70 & 0.86 \\
HD~75732 &   16262.016602 &  -3.30 & 0.87 \\
HD~75732 &   16262.040039 &  -0.83 & 0.81 \\
HD~75732 &   16262.050781 &  -2.15 & 0.85 \\
HD~75732 &   16262.053711 &  -1.37 & 0.83 \\
HD~75732 &   16262.055664 &  -0.69 & 0.78 \\
HD~75732 &   16262.057617 &  -0.72 & 0.78 \\
HD~75732 &   16262.059570 &  -2.44 & 0.77 \\
HD~75732 &   16262.061523 &   0.36 & 0.78 \\
HD~75732 &   16262.062500 &   1.05 & 0.80 \\
HD~75732 &   16262.064453 &  -1.57 & 0.78 \\
HD~75732 &   16262.066406 &   0.93 & 0.88 \\
HD~75732 &   16262.068359 &  -2.44 & 0.82 \\
HD~75732 &   16262.070312 &  -1.81 & 0.89 \\
HD~75732 &   16262.072266 &  -1.92 & 0.83 \\
HD~75732 &   16262.075195 &  -2.77 & 0.82 \\
HD~69839 &   16260.981445 &   0.08 & 0.87 \\
HD~69839 &   16260.984375 &   0.12 & 0.82 \\
HD~69839 &   16260.986328 &   0.52 & 0.87 \\
HD~69839 &   16260.989258 &   0.03 & 0.82 \\
HD~69839 &   16260.991211 &   0.73 & 0.80 \\
HD~69839 &   16260.993164 &  -1.39 & 0.82 \\
HD~69839 &   16260.995117 &   0.17 & 0.81 \\
HD~69839 &   16260.998047 &   0.35 & 0.80 \\
HD~69839 &   16261.000000 &   0.14 & 0.80 \\
HD~69839 &   16261.001953 &   0.38 & 0.87 \\
HD~72673 &   16261.040039 &   4.98 & 0.92 \\
HD~72673 &   16261.043945 &  -0.63 & 0.82 \\
HD~72673 &   16261.048828 &   1.74 & 0.90 \\
HD~72673 &   16261.053711 &  -0.54 & 0.89 \\
HD~72673 &   16261.057617 &   2.34 & 0.85 \\
HD~72673 &   16261.061523 &   0.91 & 0.98 \\
HD~72673 &   16261.065430 &   2.64 & 0.89 \\
HD~72673 &   16261.070312 &   0.39 & 0.87 \\
HD~72673 &   16261.074219 &   1.42 & 0.91 \\
HD~72673 &   16261.078125 &   1.27 & 0.93 \\
\enddata
\mylabelt{rvs}
\end{deluxetable}

We then took advantage of the stabilized SLSF. For each of the individually analyzed spectral chunks in the iodine region, we calculated a median SLSF based on all B-star observations and kept the SLSF fixed in our analysis. Because HIRES is not stabilized, this is not a perfect model; however, it is more accurate than fitting the SLSF for each observation. We used this fixed median SLSF in two critical steps. First, we adopted the median SLSF when deconvolving spectra to produce a template intrinsic stellar spectrum used in the forward modeling of our program observations. Second, we fixed the median SLSF in our analysis of the program observations, reducing the number of free parameters from 21 to 4 for each chunk.

In \tabl{fixedfree}, we compare the results of our two analysis strategies: the standard approach where we solve for the 17 SLSF parameters and the modified approach where we make use of the stability of the SLSF provided by the double scrambler and fix the SLSF to the median values from our B-star observations. For the seven stars that were observed with the double scrambler, we obtained radial velocities using both modeling algorithms (free SLSF and fixed median SLSF). We find that both the single-measurement precision and the velocity RMS improved in every case except one (HD~9407) when modeling our data with the frozen SLSF. On the other hand, velocity RMS for HD 69830 improved by more than a factor of 4. With this analysis, we improved the single measurement precision from an average of 0.99 \ms\ to 0.87 \ms. If we exclude HD~75732, which has a known planet with an orbital period of 0.7369 days, the velocity RMS improves from an average of 2.13 \ms\ to 1.48 \ms\ (see \tabl{fixedfree}). If we include HD~75732, the velocity RMS scatter of the observed stars decreases from 2.50 \ms\ to 1.64 \ms. The radial velocities obtained with the fixed SLSF algorithm are listed in \tabl{rvs}.

For completeness, we went back and reanalyzed the single circular fiber data obtained on 2010 September 30, using the median SLSF obtained from 5 B-star observations during that run. For this case, the RMS velocity scatter with the fixed, median SLSF was worse than the RMS when we fit for the SLSF with 17 free parameters. It is possible that the 5 B-star observations were insufficient for characterizing the true SLSF of the circular fiber data; fixing an incorrect SLSF into the analysis will always yield a poorer fit. The increase in the RMS velocity scatter is also consistent with the hypothesis that the single fiber scrambler did not adequately stabilize the SLSF; double scrambling is required.

\begin{deluxetable}{cccccc}
\tablecaption{Comparison of velocities with free and fixed SLSF}
\tablewidth{0pt}
\tablehead{ \colhead{ } & \colhead{ } & \colhead{Free SLSF} & \colhead{Fixed SLSF} & \colhead{Free SLSF} & \colhead{Fixed SLSF} \\
\colhead{Star} & \colhead{$\mbox{N}_{\mbox{obs}}$} & \colhead{SMP [\ms]} & \colhead{SMP [\ms]} & \colhead{RMS [\ms]} & \colhead{RMS [\ms]} \\}
\startdata
HD~9407  & 10 & 0.94 & 0.83 & 1.76 & 2.17 \\ 
HD~10700 & 44 & 1.13 & 1.1 & 2.41 & 1.68 \\ 
HD~26965 & 25 & 1.03 & 0.92 & 1.87 & 1.66 \\ 
HD~32147 & 25 & 0.84 & 0.72 & 1.87 & 1.13 \\ 
HD~69830 & 10 & 0.97 & 0.83 & 2.46 & 0.57 \\ 
HD~72673 & 10 & 1.07 & 0.89 & 2.45 & 1.65 \\ 
HD~75732 & 33 & 0.92 & 0.80 & 4.65 & 2.61 \\ 
\enddata
\mylabelt{fixedfree}
\end{deluxetable}

In order to set a baseline for the expected Doppler precision, we carried out theoretical simulations. We used the NSO solar spectral atlas with the same wavelength range as the iodine spectra. We introduced a Doppler shift, degraded the resolution, added Poisson noise to the spectra, and used a Levenberg-Marquardt algorithm to calculate the weighted mean radial velocity. At each step of SNR, the standard deviation from 10,000 Monte Carlo trials provided the theoretically expected radial velocity error as a function of spectral resolution and SNR. 

In \fig{rv}, we show the dependence of single-measurement precision on the signal-to-noise ratio of the observations. The solid lines indicate our theoretical simulations of Doppler precision as a function of SNR for two instrumental resolutions: R=55,000 (the usual Keck slit resolution) and R=70,000 (the resolution of the double scrambler). The empirical data in this plot are all measurements of the radial velocity standard star $\tau$ Ceti. The blue star-shaped symbols show the single-measurement precision as a function of SNR obtained from the double scrambler data and analyzed with the fixed median SLSF. This can be compared with the red squares, which show the single measurement precision for all $\tau$ Ceti observations made with the 0.861-arcsec B5 decker (R ~ 55,000) at HIRES between 1995 - 2012 and analyzed with free SLSF parameters. The data points for the slit data (red squares) separate into two distributions; the lower precision group of points corresponds to Doppler data obtained before the HIRES CCD upgrade in Aug 2004, and the higher precision points were taken after the CCD upgrade. The green plus signs were obtained with the B1 HIRES decker and have a resolution of R$\sim$72,000, similar to the double scrambler. The green plus signs with larger errors were obtained before the 2004 CCD upgrade.

From this plot, we see that higher resolution translates to better precision as expected.
We also see that the precision of the double scrambler data exceeds the precision of the slit data at all SNR levels, even at similar resolutions. At SNR = 350, the single measurement precision with the double scrambler is 0.75 \ms, compared to 1 \ms\ for the slit observations. Furthermore, the SMP for data obtained with the double scrambler is generally closer to theoretically obtainable SMP for a given spectrometer resolution and SNR.
It is also apparent that there is a knee in the precision floor at a SNR of about 250 with the slit-fed HIRES data. Doubling the exposure time to increase the SNR from 250 to 350 only yields a SMP gain of 0.1 \ms, which is essentially lost in systematic errors of the Doppler iodine technique.

\begin{figure}[ht]
\epsscale{0.75}
\plotone{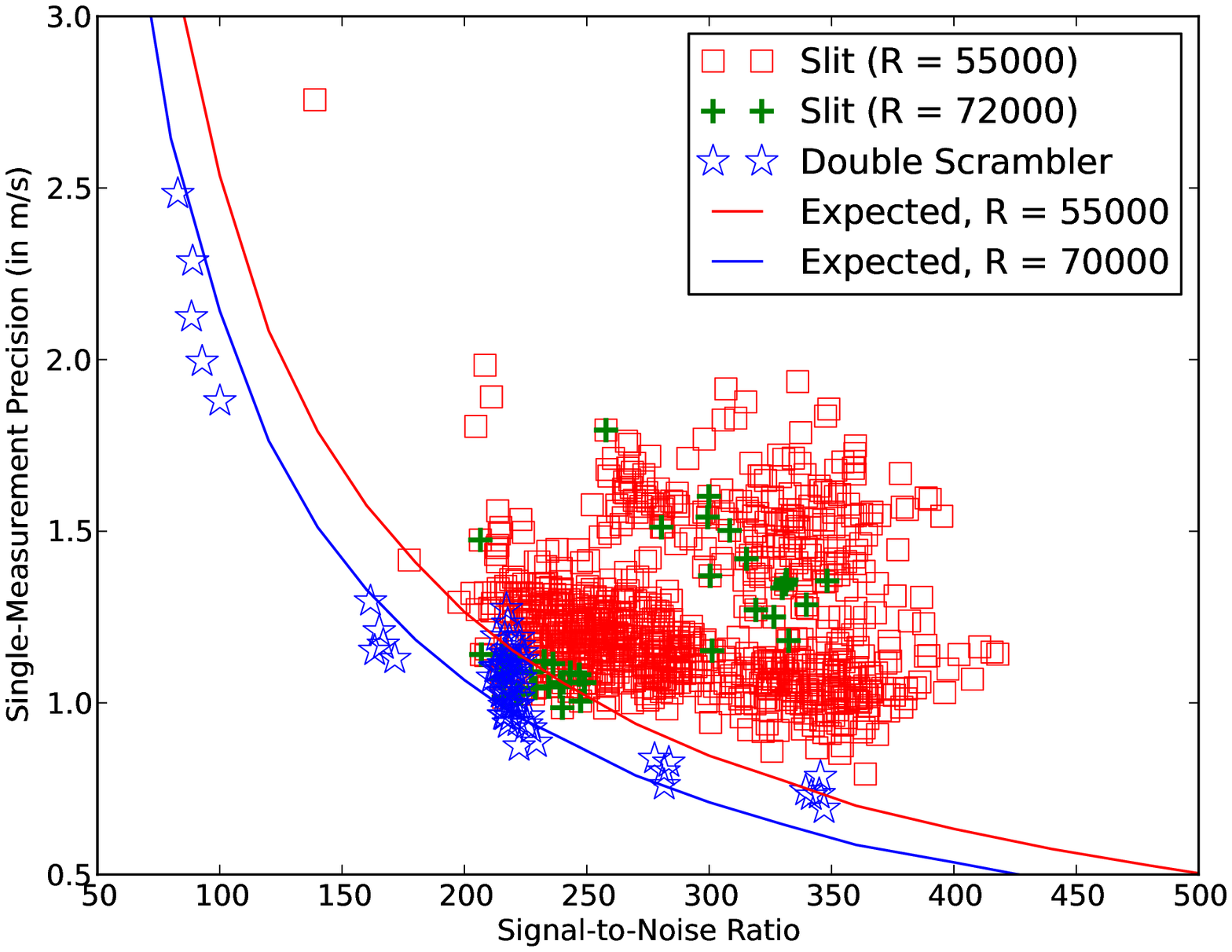}
\caption{Single-measurement precision as a function of signal-to-noise ratio (SNR). The solid curves are theoretical calculations of Doppler precision as a function of SNR and spectral resolution. All of the individual points refer to measurements of $\tau$ Ceti. The single measurement precision for observations obtained with the double scrambler are shown with blue star symbols; the historical HIRES measurements obtained with the slit are plotted as red squares. The green plus signs are high-resolution slit data.}
\mylabelf{rv}
\end{figure}

In \tabl{rvrms}, we compare the radial velocities obtained with all three methods for coupling light into HIRES: the slit, the circular fiber and the double scrambler (using the fixed median SLSF). The comparison is limited to two stars, HD~26965 and HD~32147, because these were the only data sets with similar quality for all three coupling methods. Despite a lower SNR, the double scrambler data provides both improved single-measurement precision and a radial velocity RMS scatter that was 30-40\% lower than the velocity RMS of the slit-fed data.

\begin{deluxetable}{cccccc}
\tablecaption{Comparison of Slit and Fiber Doppler Precision}
\tablewidth{0pt}
\tablehead{ \colhead{Star} & \colhead{Slit/Fiber} & \colhead{$\mbox{N}_{\mbox{obs}}$} & \colhead{SNR} & \colhead{SMP} & \colhead{RMS} \\}
\startdata
HD~26965 & Slit &  25  & $302 \pm 8$ & $0.98 \pm 0.04$ & 2.11 \\ \
HD~26965 & Circular Fiber   &  25  & $259 \pm 3$ & $1.37 \pm 0.05$ & 2.46 \\ \
HD~26965 & Double Scrambler & 50  & $217 \pm 1$ & $0.92 \pm 0.05$ & 1.48 \\ \
HD~32147 & Slit       & 28   & $288 \pm 20$ & $2.10 \pm 0.07$ & 1.96 \\ \
HD~32147 & Circular Fiber   & 34   & $265 \pm 1$ & $2.40 \pm 0.08$ & 1.95 \\ \
HD~32147 & Double Scrambler & 31  & $224 \pm 8$ & $0.73 \pm 0.11$ & 1.49 \\ 
\enddata
\mylabelt{rvrms}
\end{deluxetable}

\begin{figure}[ht]
\epsscale{0.75}
\plotone{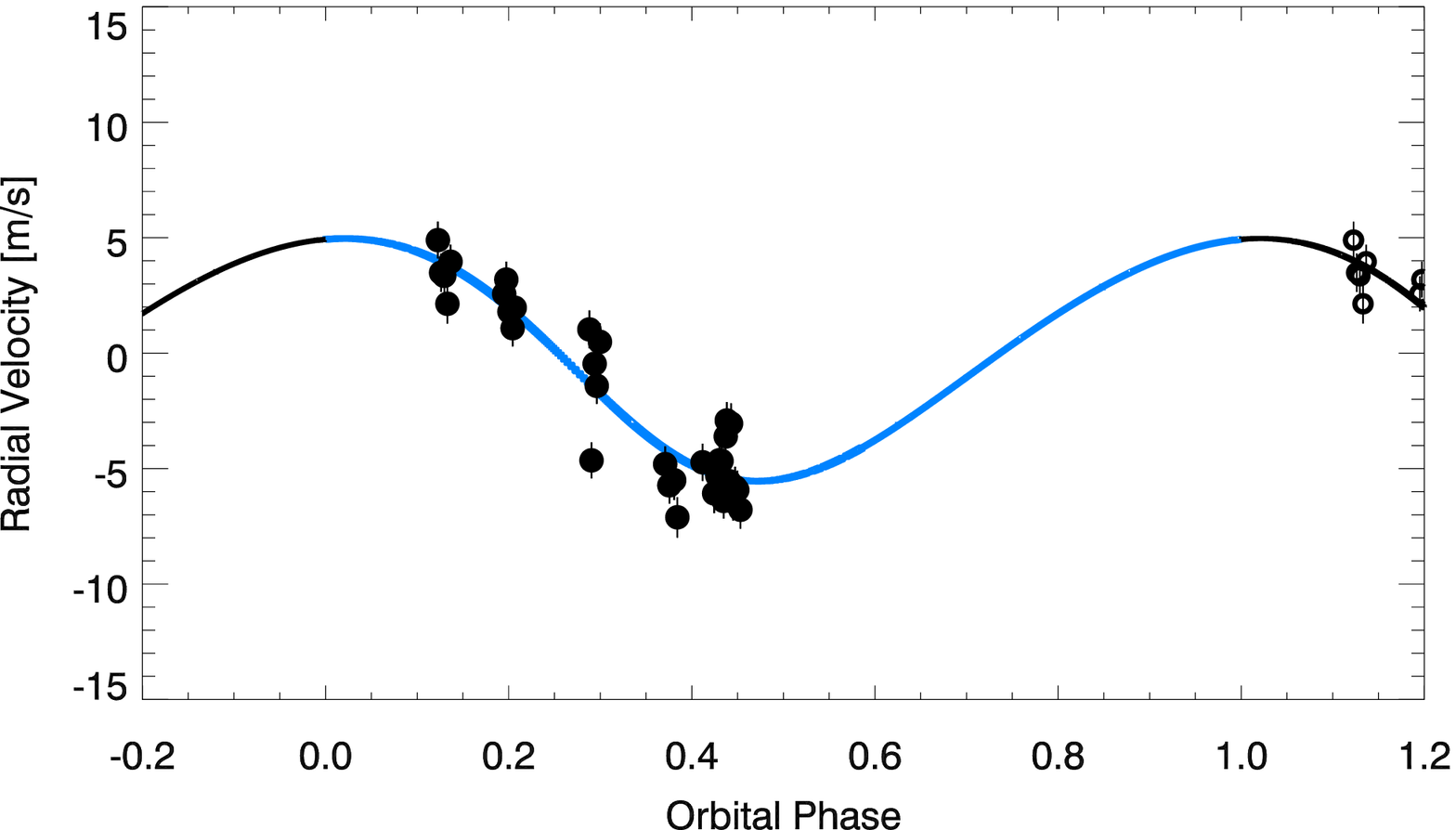}
\caption{Radial velocities of 55 Cnc e obtained with the double scrambler are phase-folded and fitted with a Keplerian model with an orbital period of 0.7369d and eccentricity fixed to 0. Points for repeated phase are shown as circles instead of dots. The RMS of the residual velocities is 1.42 \ms.}
\mylabelf{55cnc}
\end{figure}

We obtained Doppler measurements of 55 Cnc throughout both nights with the double fiber scrambler. Although the phase coverage for the orbit is incomplete, we fitted the data using the IDL program Keplerian Fitting Made Easy \citep{Giguere2012} using an initial guess for the orbital period of 0.737 d and an orbital eccentricity fixed to 0. A bootstrap Monte Carlo was run with 400 trials to return the parameter uncertainties listed in \tabl{55cnc}; however, the incomplete phase coverage probably led to an underestimate of the uncertainty in the orbital parameters. After removing the best fit Keplerian curve, the residual velocity RMS is 1.42 \ms. 

\begin{deluxetable}{llc}
\tablecaption{Keplerian Model Parameters for 55 Cnc e}
\tablewidth{0pt}
\tablehead{ \colhead{Parameter} & \colhead{Value} & \colhead{Comments} \\}
\startdata
$P$        & 0.7369 $\pm 0.1215$ d   &         \\ 
$ecc$      & 0                       & (fixed) \\ 
$Tp$       & 2456262.08 $\pm 0.2$ d  & BJD     \\ 
$K$        & 4.54 $\pm 0.5$ \ms\     &         \\ 
$M \sin i$ & 6.4 $M_\oplus$          &         \\ 
RMS        & 1.47 \ms\               &         \\ 
\enddata
\mylabelt{55cnc}
\end{deluxetable}

\section{Conclusions}
We tested a prototype double scrambler for
Keck HIRES that improves spectral line spread function (SLSF) stability by an order of magnitude compared to a single fiber scrambler and uses a pupil slicer to increase spectral resolution to R = 70,000 with minimal slit losses.. With this design, we inject starlight from the Keck I telescope into a 200-micron octagonal fiber. The pupil after this fiber is then sliced in two half-moons and focused onto the front face of a $100 \times 400$-micron rectangular fiber. The output of the second fiber replaced the entrance slit of the HIRES spectrograph for a test run of this prototype. 

We observed 56 B-stars throughout the two nights of 2012 November 28 and 29. These observations provide a measure of the spectral line spread function (SLSF). To quantify SLSF stability, we fitted each SLSF with a Gaussian and then calculated the standard deviation of the FWHM for all comparable observations. The SLSF for the double scrambler was remarkably stable; the standard deviation of FWHM was only 0.018 pixels. Using the same stability metric, the double scrambler was 9 times more stable than a single fiber scrambler we tested in 2010 and 18 times more stable than a slit with no fiber. Thus, the double fiber scrambler met our goal of improving scrambling to produce a more stable SLSF. One weakness of our prototype is that the throughput was lower at the telescope than in the lab. Better throughput must be achieved before the double scrambler will be practical for general use with high-resolution spectrographs.

Over two nights, we obtained a series of spectra for six radial velocity standard stars and successfully detected the 0.7369-day signal from 55 Cnc e. We analyzed the data in two ways. First, we carried out a Doppler analysis that was identical to what we would do for observations without a fiber. With this analysis, we did not see any improvement in the radial velocity precision, despite the fact that the SLSF was dramatically more stable. In our second analysis of the same data, we took advantage of the stabilized SLSF and calculated a median SLSF based on all 56 B-star observations. This median SLSF was fixed in both the template deconvolution and in the forward modeling of our program observations with the iodine cell. Because HIRES is not an environmentally stabilized instrument, the median SLSF is still imperfect and there are additional modeling errors from inaccurate deconvolution of both the template and the FTS iodine scan. However, using the double scrambler and the fixed-SLSF algorithm, we improved the RMS scatter from an average of 2.1 \ms\ to 1.5 \ms. 

We find that the error contributed by inaccurate modeling of the SLSF is $\sim$1.47 \ms\ for slit-fed HIRES (the quadrature sum of 1.47 and 1.5 is 2.1 \ms).  The information in our spectral chunks is insufficient to constrain the SLSF model while also solving for the wavelength and Doppler shifts. These modeling errors do not average down like Poisson noise. Despite single measurement errors of about 1 \ms, published results for chromospherically quiet stars still have residual velocity RMS greater than 1.5 \ms\ and hundreds of observations are required to fit a 6-parameter Keplerian model for low amplitude exoplanets. 

This work also shows that stellar activity does not set the floor of the Doppler precision for chromospherically inactive stars observed with Keck HIRES and the iodine reference cell. Analyzing the contribution of stellar jitter to radial velocity uncertainties, \citet{Isaacson2010} found that the velocity jitter was independent of stellar activity for K dwarfs, leading them to suggest that stellar jitter for K dwarfs was simply below a $\sim$1.6 \ms\ instrumental precision floor of CPS Doppler program. This conclusion is consistent with our findings here. 

This work sheds light on how to move toward 10 \cms\ Doppler precision. Significantly higher resolution spectrometers with extreme environmental stability, uniform illumination and precise wavelength calibrators are critical. Expanding the wavelength range beyond the iodine region will provide more leverage for higher precision. Perhaps most important, the dynamical motion of the star must be distinguished from stellar photospheric signals. Eliminating iodine as a wavelength calibrator also has the advantage of providing a clearer view of photospheric velocities embedded in the spectral line profiles. Much higher resolution should also help to identify photospheric signals. These are the next big challenges to address for Doppler planet searches.

\acknowledgments
We thank Tom Blake for help obtaining the FTS scan of the iodine cell at EMSL, a DOE Office of Science User Facility sponsored by the Office of Biological and Environmental Research and located at Pacific Northwestern National Labs.We acknowledge support from the Planetary Society, who made possible the development of fiber scramblers at Lick and Keck Observatories. We thank the support teams at Keck for their help with installation and implementation of the fiber scramblers, particularly Scott Dahm and Grant Hill. DAF and JFPS acknowledge research support from NSF grant AST-1207748 and NASA grant NNX12AC01G. The authors wish to acknowledge the very significant cultural role and reverence that the summit of Mauna Kea has within the indigenous Hawaiian community. We are most grateful for the opportunity to conduct observations from this mountain.


\bibliographystyle{plainnat}

\end{document}